\documentclass[submission, Phys]{SciPost}
\usepackage[normalem]{ulem}

\usepackage{graphicx}
\usepackage{amsmath}
\usepackage{amssymb}
\usepackage{psfrag,epsfig}
\usepackage[percent]{overpic}
\usepackage{float}
\usepackage{color}
\usepackage{hyperref}

\begin{document}

\begin{center}{\Large \textbf{Universal Chern number statistics in random matrix fields}}
\end{center}

\begin{center}
Or Swartzberg\textsuperscript{1*}, Michael Wilkinson\textsuperscript{2,3$\dagger$}, and Omri Gat\textsuperscript{1$\ddag$}
\end{center}

\begin{center}
{\bf 1} Racah institute of Physics, Hebrew University, Jerusalem 91904, Israel
\\
{\bf 2} Chan Zuckerberg Biohub, 499 Illinois Street, San Francisco, CA 94158, USA
\\
{\bf 3} School of Mathematics and Statistics, The Open University, Walton Hall, Milton Keynes, MK7 6AA, England
\\
*or.swartzberg@mail.huji.ac.il
$^\dagger$ m.wilkinson@open.ac.uk
$^\ddag$ omrigat@mail.huji.ac.il
\end{center}

\begin{center}
\today
\end{center}

% For convenience during refereeing: line numbers
%\linenumbers

\section*{Abstract}
{\bf
We investigate the probability distribution of Chern numbers (quantum Hall conductance integers) 
for a parametric version of the GUE random matrix ensemble, which is a model for a chaotic or 
disordered system. The numerically-calculated single-band Chern number statistics agree well with 
predictions based on an earlier study 
[O.\ Gat and M.\ Wilkinson, \textit{SciPost Phys.}, {\bf 10}, 149, (2021)] 
of the statistics of the quantum adiabatic curvature, when the parametric correlation length is small. 
However, contrary to an earlier conjecture, we find that the gap Chern numbers are correlated, and 
that the correlation is weak but slowly-decaying. Also, the statistics of weighted sums of Chern 
numbers differs markedly from predictions based upon the hypothesis that gap Chern numbers
are uncorrelated. All our results are consistent with the universality hypothesis described in the 
earlier paper, including in the 
previously unstudied regime of large correlation length, where the Chern statistics is 
highly non-Gaussian.}

\vspace{10pt}
\noindent\rule{\textwidth}{1pt}
\tableofcontents\thispagestyle{fancy}
\noindent\rule{\textwidth}{1pt}
\vspace{10pt}

\section{Introduction}
\label{sec: 1}

The discovery of the quantum Hall effect \cite{vKl+80} was 
followed by an observation that the Hall conductance integer $N_n$
for systems with a band spectrum can be expressed as 
a Chern integer \cite{Tho+82,Tho83}. This is a topological invariant that can be expressed 
as an integral of {a curvature $\Omega_n$} 
over a closed surface ${\cal S}$ \cite{Che+74}:
\begin{equation}
\label{eq: 1.1}
N_n=\frac{1}{2\pi}
\int_{\cal S}{\rm d}\mbox{\boldmath$X$}\ \Omega_n(\mbox{\boldmath$X$})
\ .
\end{equation}
Here $n$ is an index labelling an isolated band of the spectrum. In the context 
of studies of the quantised Hall effect, the curvature is the quantum adiabatic curvature 
\cite{Ber84,Sim83} and the closed surface is a Brillouin zone \cite{Tho+82}, or a manifold 
specifying a boundary condition defined by complex phase factors \cite{Tho83}. 
The Hall conductance is argued to be equal to $e^2/h$ times the sum of the 
Chern integers $N_n$ for all of the filled bands. 
The discovery of the role of Chern integers has stimulated the identification of other 
topological invariants describing transport phenomena, \cite{km,scz}, 
which in turn led to a comprehensive theory of energy band topology 
and the role that symmetries therein, \cite{kitaev,ludwig}. A key feature of 
topologically nontrivial energy bands is the appearance of edge states that are robust 
under perturbations \cite{hasan}. Majorana modes are a class of edge states that has 
attracted a lot of interest because of possible applications to quantum 
computing \cite{majorana}.

It is instructive to assess whether different approaches to understanding 
a phenomenon lead to consistent predictions. In the case of the Hall conductance, 
we can ask whether the Hall conductance expressed in terms of Chern 
integers is compatible with estimates that can be obtained by semiclassical 
approaches, such as those appearing in traditional texts on solid-state physics \cite{Zim64}. 
(These might be based upon the Boltzmann equation, or upon making a classical 
estimate for the correlation function appearing in the Kubo formula.)  
To state this question more formally, we can consider a limit where the size 
of the unit cell is increased, leaving the Fermi energy and other parameters 
constant. In this limit the number of bands below the Fermi energy increases. 
In order to compare the semiclassical estimate with the Hall conductance 
which is obtained from the Chern integers, it is necessary to determine the properties of the integers 
$N_n$, and in particular their typical magnitude, when the number of bands is large. 
Moreover, because the Hall conductance integer is the 
sum of the values of $N_n$ for the filled bands, correlations between different values 
of $N_n$ are of interest: if the $N_n$ were uncorrelated random numbers, this might imply 
very large values of the Hall conductance. We might, therefore, anticipate that the 
there are correlations of the $N_n$ which reduce the magnitude of their sum.

Building on seminal work by Wigner \cite{Wig55} and Dyson \cite{Dys62}, 
it is now understood that the spectra 
of complex quantum systems resemble those of samples from random matrix ensembles \cite{Meh91}. 
For our purposes, complex quantum systems are defined as those having many levels 
and no symmetries, localisation effects, or constants of motion.
The Chern integers, however, are defined for families of Hamiltonians depending 
upon parameters---Bloch Hamiltonians depending on lattice momenta for the quantum Hall
conductance. The random matrix theory approach has been extended to study correlations 
of energy levels and wave function of parameter-dependent complex systems 
\cite{Wil88,Wil89,Aus+92,SiAl93,SzAl93,BeRe93,AtAl95}. 

A good starting point to understand the Chern numbers in complex systems is therefore to analyse the 
Chern numbers for parameter-dependent random matrix models. 
The goal of the present work is to examine the scaling, correlations, and universality 
of Chern numbers in parametric Gaussian Unitary Ensemble (GUE) models using 
Monte-Carlo calculations. {Our work builds upon some earlier investigations 
which have applied parametric random matrix theory to the quantum adiabatic 
curvature and Chern integers. In particular, \cite{Wal+95} addressed the statistics of 
Chern integers by calculating the statistics of degeneracies in a three-parameter 
parametric random matrix model: the results are consistent with the more refined 
investigation presented here. The single-point probability distribution 
of the curvature was obtained for a parametric random matrix model 
in \cite{Ber+18}. The curvature in complex systems has also 
been studied using semiclassical approximations in \cite{Ber+92,Robbins94}.}    

A recent paper, {\cite{Gat+21}}, by two of the present authors, referred to as paper I in the following, 
studied the statistics of the curvature, $\Omega_n(\mbox{\boldmath$X$})$, for parametric GUE models, 
and the results were 
used to propose expressions for the variance and the correlations of the Chern numbers. 
The comparison with the formulae proposed 
in paper I is informative: the prediction of the variance of the Chern integers 
proved to be very accurate, but there are significant deviations from the expression 
which was proposed for the correlations of weighted sums of Chern integers of different bands.
Namely, the prediction of paper I regarding multiband Chern number correlations was based 
on the hypothesis that the gap Chern numbers (that is, cumulative sums of band Chern numbers) 
are statistically independent. Our first main result is to show that this hypothesis is false: the 
correlations between the gap Chern numbers are small, but they decay slowly as a power law for 
increasing the spectral separation. 

Our second main results concerns the band Chern number statistics when the parametric 
correlation length, equal to the inverse of the product of the density of states and the parametric 
sensitivity of the random matrix elements {\cite{Gat+21}}, is not small. When the parametric correlation 
length is taken to zero, the Chern number probability distribution is approximately Gaussian, and its 
variance tends to a constant limit when properly scaled. When the correlation length is not small, on the 
other hand, the distribution is far from Gaussian, and its moments depend in a complicated way on the 
correlation length. Nevertheless, we find this dependence becomes \emph{universal} as the matrix size increases.

The paper is organised as follows. 
{Section \ref{sec: 2} defines the 
parametric GUE model to be studied. 
Section \ref{sec: 3} reviews the results and hypotheses regarding the statistics 
of the Chern integers which were developed 
in paper I. Section \ref{sec: 4} presents our results for the statistics of the Chern number of a
single band.
In section \ref{sec: 5} we analyze the correlations of gap Chern numbers and their dependence 
on the spectral separation between the two gaps.
In section \ref{sec: 6} we study the correlations of weighted sums of band Chern numbers, 
and their deviation from the predictions based on the results of paper I (arising from 
multi-level Chern number correlations). Section \ref{sec: 7} offers our conclusions from this work.

\section{Parametric GUE models}
\label{sec: 2}

To study the statistics of the Chern numbers, we used a model for fields of random matrices 
constructed using an approach considered in \cite{Wil89,AtAl95}. {We define an ensemble 
of fields of random matrices {$H(\mbox{\boldmath$p$})$} over a parameter space $\mathcal{S}$. The 
ensemble is statistically homogeneous and isotropic in the parameter space, and it} exhibits the following properties:
 
\begin{enumerate}

\item For any $\mbox{\boldmath$p$}\in \mathcal{S}$, the matrix $H(\mbox{\boldmath$p$})$ is 
a representative of the Gaussian Unitary 
Ensemble (GUE) of complex Hermitean random matrices, as defined in \cite{Dys62,Meh91}. 
The matrices have dimension {$M$}. The elements are independently Gaussian distributed, with zero mean, 
and their magnitude has unit variance. 

\item The matrix element correlations are
\begin{align}
 \label{eq: 2.1} 
\langle H_{ij}{(\mbox{\boldmath$p$}_1)^*H_{ij}(\mbox{\boldmath$p$}_2)\rangle=
c(|\mbox{\boldmath$p$}_1-\mbox{\boldmath$p$}_2|)}
\end{align} 
for some correlation function $c$ {satisfying $c(0)=1$ and positivity}, and different matrix 
elements (other than those related by Hermiticty) 
are uncorrelated at different points in the parameter space.
\end{enumerate}

In our work we took the manifold $\mathcal{S}$ to be the unit sphere embedded in three-space, 
and the distance between points {$\mbox{\boldmath$p$}_1$, $\mbox{\boldmath$p$}_2$} on the 
sphere is taken as the Euclidean distance. 
Our numerical investigations included three different one-parameter families of correlation functions:

\begin{enumerate}

\item A Gaussian correlation function: $c(|{\mbox{\boldmath$p$}_1-\mbox{\boldmath$p$}_2|)
=\exp(|\mbox{\boldmath$p$}_1-\mbox{\boldmath$p$}_2}|^2/(2r^2))$.

\item  A Lorenzian correlation function: 
$c(|{\mbox{\boldmath$p$}_1-\mbox{\boldmath$p$}_2|)
=1/(1+(|\mbox{\boldmath$p$}_1-\mbox{\boldmath$p$}_2|/l)^2)}$.

\item A {\emph{four-matrix model}} of the form
\begin{equation}
\label{eq: 2.3}
{H(\mbox{\boldmath$r$})={\cos\alpha H_0+\sin\alpha \,\mbox{\boldmath$r$}\cdot H_{\mbox{\boldmath$r$}}}}
\end{equation}
where $H_0$ is a GUE matrix, {$H_{\mbox{\boldmath$r$}}$ is a vector of three GUE matrices and 
$\mbox{\boldmath$r$}$} is a unit vector on the sphere, implying that
\begin{equation}
\label{eq: 2.4}
{c(|\mbox{\boldmath$r$}_1-\mbox{\boldmath$r$}_2|)
=\cos(\alpha)^2+\sin(\alpha)^2
\frac{2-|\mbox{\boldmath$r$}_1-\mbox{\boldmath$r$}_2|^2}{2}}
\ .
\end{equation}
\end{enumerate}

According to the principles discussed in \cite{Wil88,Wil89,Aus+92}, the  universal
properties of this parametric random matrix model depend upon the density of states, $\rho$, 
and upon a tensor describing the sensitivity of the energy levels to displacements in the parameter 
space. For an isotropic model, this tensor is a represented by a diagonal matrix, and the 
parametric sensitivity is described by a single parameter, $\sigma^2$, which is the variance 
of first derivative of the levels:
\begin{equation}
\label{eq: 2.2}
\sigma^2={\rm var}\left(\frac{\partial E_n}{\partial p}\right)=-\frac{1}{2}\frac{{\rm d}^2c}{{\rm d}p^2}\bigg\vert_{p=0}
\ .
\end{equation}
The parametric sensitivity can be any positive number in the Gaussian and Lorentzian families, and is 
between zero and one in the four-matrix model. In paper I there is a discussion of how a linear 
transformation of the parameter space {of this model} can be used to model systems in which the parameter 
space is not isotropic.

The spectral correlations of the parametric GUE model \eqref{eq: 2.1} were previously 
studied in \cite{AtAl95}. Here we study the Chern number statistics numerically by 
sampling matrices of the parametric GUE distribution on the vertices of a triangulation 
of the two-sphere. For each such 
realisation we calculated the Chern number using 
the algorithm of \cite{Fuk+05}, refining the triangulation until convergence was obtained.
In this manner we calculate the distribution of the Chern numbers for the three 
families of matrix-element correlations listed above, with different matrix sizes and 
correlation lengths. The details of the numerical method are described in appendix \ref{app:mc}.

\section{{Theoretical background}}
\label{sec: 3}

The correlation function of the curvature field $\Omega_n(\mbox{\boldmath$p$})$ was studied in \cite{Gat+21}.
It was argued that when the matrix size is large this correlation function may be expressed in the scaling form
\begin{equation}
\label{eq: 3.1}
\langle \Omega_n(p_1)\Omega_n(p_2)\rangle=(\rho\sigma)^4f(\rho\sigma 
|{\mbox{\boldmath$p$}_1-\mbox{\boldmath$p$}_2}|) 
\end{equation}
where $f(x)$ is a universal
function, which approaches zero rapidly for large $|x|$. 
Equation \eqref{eq: 3.1} implies that the curvatures at points in parameter 
separated by a distance much larger than ${1}/({\rho\sigma})$ are uncorrelated. It follows, by
taking the square of (\ref{eq: 1.1}) and averaging, that when $\rho\sigma\gg1$
\begin{equation}
\label{eq: 3.2}
\langle N_n^2\rangle=\frac{1}{2\pi}{\cal A}\sigma^2\rho^2{\cal I}
\end{equation}
where ${\cal A}$ is the area of the surface $\mathcal{S}$ and ${\cal I}\approx 1.69$
is a universal constant obtained by integrating the function $f$. We note that $f(x)$ diverges 
as $1/x$ when $x$ tends to zero because the curvature distribution has infinite variance \cite{Ber+18}, 
but the integral of $f$ over the two-dimensional parameter space is finite.

Reference \cite{Gat+21} also discussed statistics of a weighted average of $\Omega_n(\mbox{\boldmath$p$})$, 
defined by 
\begin{eqnarray}
\label{eq: 3.3}
\bar \Omega_\epsilon(E,\mbox{\boldmath$p$})&=&\sum_n\Omega_n(\mbox{\boldmath$p$})\delta_\epsilon(E-E_n)
\nonumber \\
\delta_\epsilon(E)&=&\frac{1}{\sqrt{2\pi}\epsilon}\exp(-E^2/2\epsilon^2)
\ .
\end{eqnarray}

The correlation function 
\begin{equation}
\label{eq: 3.4}
{\cal C}_\epsilon(\Delta E,\Delta \mbox{\boldmath$p$})\equiv
\langle\bar \Omega_\epsilon(E+\Delta E,\mbox{\boldmath$p$}+\Delta\mbox{\boldmath$p$})
\bar \Omega_\epsilon(E,\mbox{\boldmath$p$})\rangle
\end{equation}
was found in paper I to take the scaling form
\begin{equation}
\label{eq: 3.5}
{\cal C}_\epsilon(\Delta E,\Delta \mbox{\boldmath$p$})=\frac{\pi^{3/2}}{6}\frac{\sigma^4\rho^3}{\epsilon^3}
g(\rho\sigma|\Delta \mbox{\boldmath$p$}|,\Delta E/\epsilon)
\end{equation}
{in the limit $M\to \infty$,}
with $g$ a universal function of two variables, which was calculated exactly in the case where {$\Delta \mbox{\boldmath$p$}={\bf 0}$.

The fact that ${\cal C}_\epsilon(\Delta E,\Delta \mbox{\boldmath$p$})$ approaches zero rapidly as $\epsilon$ increases 
indicates a high degree of cancellation of the values of $\Omega_n(\mbox{\boldmath$X$})$ associated with different bands.
In paper I it was argued that
the fact that ${\cal C}_\epsilon\sim \epsilon^{-3}$ is consistent with the 
Chern integers having a two-point correlation of the form
\begin{equation}
\label{eq: 3.7}
\langle N_nN_m\rangle-\langle N_n\rangle\langle N_m\rangle
=\frac{1}{2}\mathop{\rm var}(N_n)\left[2\delta_{nm}-\delta_{n,m+1}-\delta_{n,m-1}\right]
\ .
\end{equation}
We can describe this relation more succinctly with the help of the
gap Chern numbers $G_n$, defined by
\begin{equation}
\label{eq: 3.8}
G_n=\sum_{i=1}^n N_i\ ,\qquad 0\le n\le N\ ,
\end{equation}
so that $N_n=G_n-G_{n-1}$, and therefore (\ref{eq: 3.7}) is consistent 
with the hypothesis that the gap Chern numbers are uncorrelated:
\begin{equation}
\label{eq: 3.9}
\langle G_nG_m\rangle=2\mathop{\rm var}(N_n)\delta_{nm}\ .
\end{equation}
In paper I this hypothesis was applied to the statistics of the weighted sum of 
Chern numbers in an energy window of width $\epsilon$, 
\begin{equation}
\label{eq: 3.6}
\bar N_\epsilon(E)=\sum_n N_n \delta_\epsilon(E-\langle E_n\rangle)\ .
\end{equation}
where $\langle E_n\rangle$ is an average of $E_n(\mbox{\boldmath$p$})$ over the parameter space.
We choose $\epsilon$ much larger than the mean level spacing, but small enough that the density of 
states does not change {appreciably} inside the energy {window. 
A direct estimate of the variance of (\ref{eq: 3.6}) involves integrating the correlation function (\ref{eq: 3.5}) 
over $\Delta \mbox{\boldmath$p$}$, and leads to the prediction that $\langle \bar N_\epsilon^2\rangle\sim \epsilon^{-3}$. 
However, the function $g(\Delta X,\Delta E)$ was not sufficiently well characterised in paper I to allow this 
integral to be estimated reliably. An  
alternative approach estimates $\langle \bar N_\epsilon^2\rangle$ using 
(\ref{eq: 3.2}) and the independent gap Chern number hypothesis, (\ref{eq: 3.7}) yielding}, 
\begin{equation}
\label{eq: 4.2}
\langle \bar N_\epsilon^2 \rangle \sim \frac{\langle N^2\rangle}{2\pi \epsilon^2}F(X)
\end{equation}
where $X=1/(2\epsilon^2 \rho^2)$, and 
\begin{equation}
\label{eq: 4.3}
F(X)=\sum_{n=-\infty}^\infty \exp(-Xn^2)[\exp(-Xn^2)-\exp(-X(n+1)^2)/2-\exp(-X(n-1)^2)/2]
\ .
\end{equation}
As explained in paper I, the expressions \eqref{eq: 4.2}, (\ref{eq: 4.3}) can be further simplified
when $\epsilon \rho\gg 1$ to obtain
\begin{equation}
\label{eq: 4.4}
\langle \bar N_\epsilon^2\rangle \approx \frac{3{\cal I}}{128\sqrt{\pi}}\frac{{\cal A}\rho\sigma^2}{\epsilon^3}\ ,
\end{equation}
exhibiting the the same $\epsilon^{-3}$ as in \eqref{eq: 3.5}. Our numerical results 
are consistent with the prediction $\langle \bar N_\epsilon^2\rangle \sim \epsilon^{-3}$, 
over a small range of $\epsilon$. However, it is found that the coefficient multiplying 
$\epsilon^{-3}$ in \eqref{eq: 4.4} is not correct.

\begin{figure}[t]
\centering
\includegraphics[width=12.5cm]{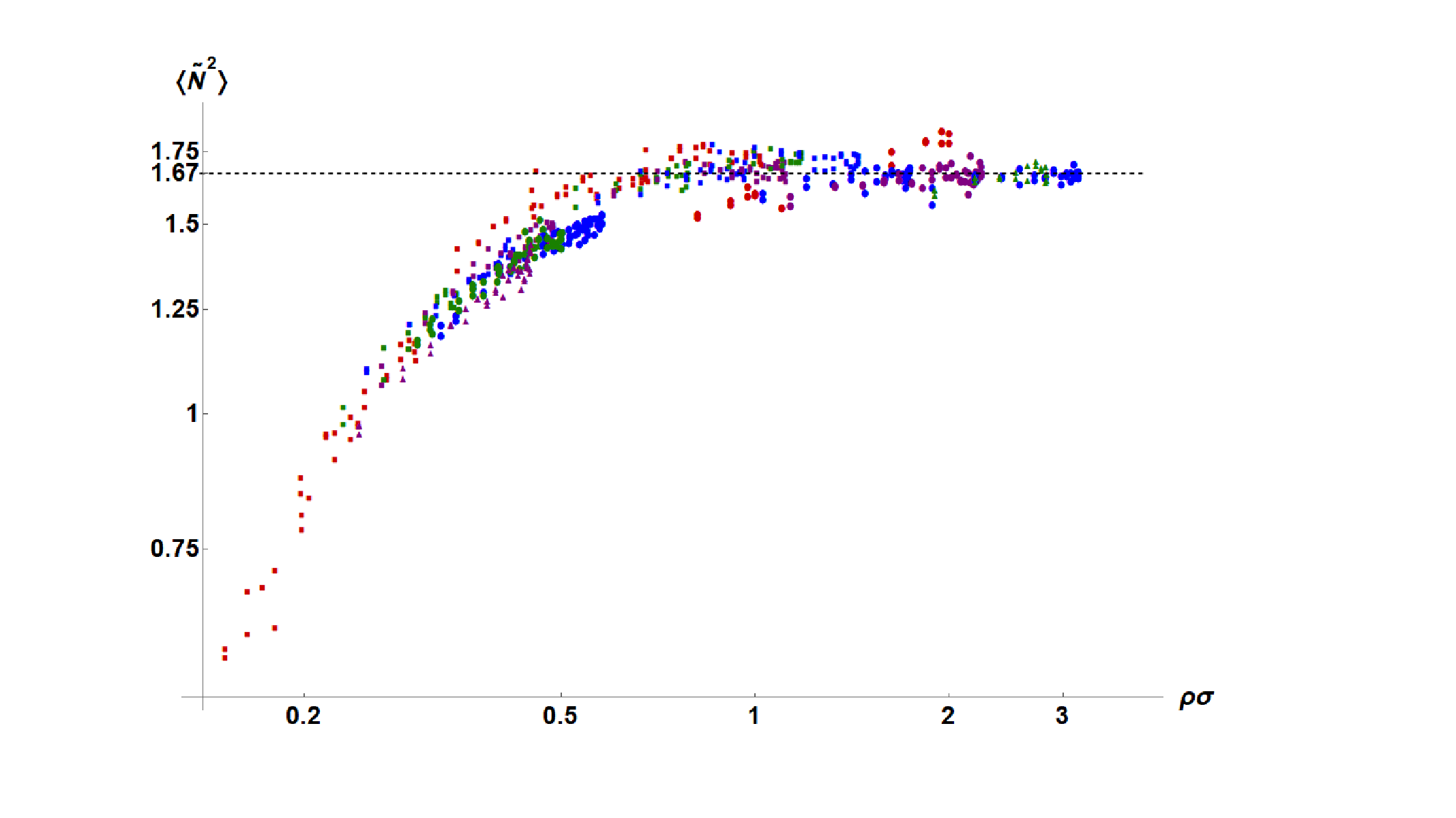}
\caption{
\label{fig: 1} Double logarithmic plot of the variance of Chern numbers of the bands in the bulk of the 
spectrum of random matrix field models, with different sizes from {$4\times 4$ to $50\times 50$}, and 
matrix element correlation functions, calculated in Monte-Carlo simulation as a function of the 
inverse parametric correlation length $\rho\sigma$. The {scaled} Chern number variance 
$\langle\tilde N_n^2\rangle{\equiv }2\pi \langle N_n^2\rangle/\rho^2\sigma^2{\cal A}$ 
approaches the universal constant $\mathcal{I}\approx1.67$ when $\rho\sigma$ is large. 
Symbol shapes correspond to Gaussian (circles), Lorentzian (triangles) and four-matrix (square) 
matrix-element correlations, and {colours to matrix dimensions $M$ according to the key: purple --- $60>M\ge 40$;
blue --- $40>M\ge30$;
green --- $30>M\ge20$;
red --- $20>M$.
}. The data collapse shows 
that band Chern number variance is a universal function of  $\rho\sigma$, {even when this variable 
takes small values, due to the scale factor ($r$ or $l$ in the Gaussian and Lorentzian random-matrix models respectively)
being large}.}
\end{figure}
\section{Band Chern number statistics}
\label{sec: 4}

We calculated the variance of Chern number {for a large number of realisations of the} random-matrix fields on the two-sphere, with {dimensions between $M=4$ and $M=50$,} with elements correlated according to the three families listed in section \ref{sec: 2}, {and} 
with several choices of scale {factors, $r$ or $l$}. Figure \ref{fig: 1} shows the {scaled variance of the} Chern integers, 
\begin{equation}
\langle\tilde N_n^2\rangle=\frac{2\pi \langle N_n^2\rangle}{\rho^2\sigma^2{\cal A}}\ ,
\end{equation}
plotted as a function of $\rho\sigma$. As expected, when $\rho\sigma\gg1$,
$\langle\tilde N_n^2\rangle$ approaches an asymptotic universal constant value ${\cal I}\approx 1.67$, 
irrespective of the shape of the correlation function, consistent with \eqref{eq: 3.2} and the estimate of 
${\cal I}\approx1.69$ based on the results of \cite{Gat+21}. We note however that the asymptotic regime 
of `large' $\rho\sigma$---small correlation length---starts already in when $\rho\sigma\gtrsim0.7$, so that the 
estimate that the parametric correlation length $\sim1/(\rho\sigma)$ involves a small proportionality factor. 
This conclusion is also consistent with the finding of \cite{Gat+21} that $xf(x)$, where $f(x)$ is the scaling 
function of \eqref{eq: 3.1} drops to half its maximal value already for $x\approx0.15$.

In the regime of large $\rho\sigma$ we furthermore expect that the Chern-number distribution is 
Gaussian because when the parametric correlation length is small, the integral in \eqref{eq: 1.1} can be 
viewed as a sum of many {nearly independent random variables, obtained by dividing the surface 
$\mathcal{S}$ into patches whose size is large compared to the parametric correlation length, $1/\rho\sigma$, 
but small compared to the entire surface}. 

\begin{figure}[tp]
\centering
\includegraphics[width=0.70\textwidth]{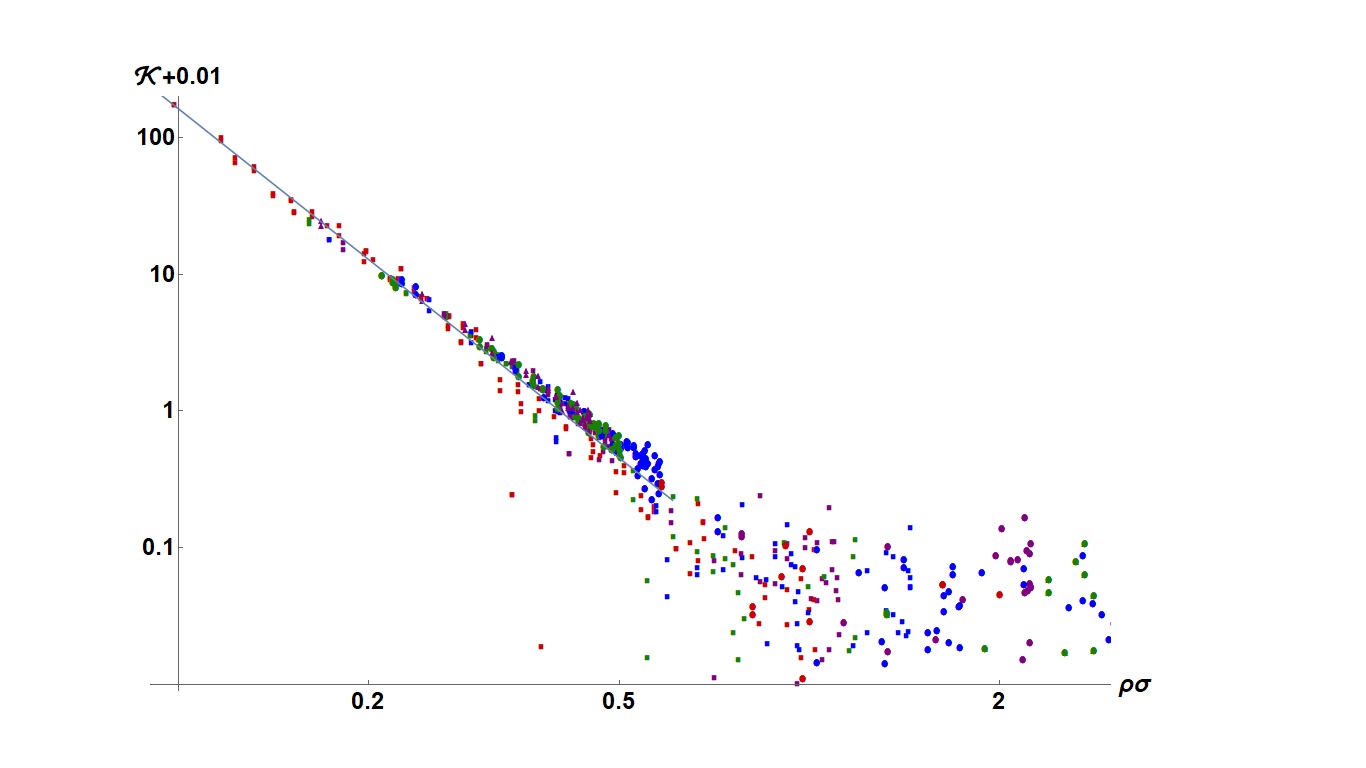}
 \caption{
 \label{fig: 11}
 Double logarithmic plot of the (excess) kurtosis $\kappa={\langle N_n^4\rangle}/{\langle N_n^2\rangle^2}-3$ 
 of the Chern number distributions whose variance is shown in figure \ref{fig: 1}, as a function 
 of $\rho\sigma$. The statistical uncertainty in the data is about 0.1, so that $\kappa$ values 
 for $\rho\sigma\gtrsim0.7$ are consistent with zero, consistent with a Gaussian Chern 
 number distribution in this regime. Note that a constant shift of $0.01$ has been added to all 
 values of $\kappa$ to improve the clarity of the graph.
The positive $\kappa$ values obtained for $\rho\sigma\lesssim0.7$ show that the Chern number distribution is 
non-Gaussian in this regime, while the data collapse further supports the hypothesis that the distribution 
is universal for all $\rho\sigma$. The straight line is a power law fit $\kappa\sim (\ell_\kappa\rho\sigma)^{-\alpha}$,
$\alpha=3.66\pm0.05$, $\ell_\kappa=2.49\pm0.05$ for $\rho\sigma\le0.7$.}
\end{figure}

We studied the convergence of the Chern-number distributions to Gaussian by calculating the (excess) kurtosis
\begin{equation}
\kappa=\frac{\langle N_n^4\rangle}{\langle N_n^2\rangle^2}-3
\end{equation}
of the randomly generated Chern-number populations; we use the definition which 
makes $\kappa=0$ in Gaussian distributions. 

Our Monte-Carlo calculations of the kurtosis are shown on a double-logarithmic plot in 
figure \ref{fig: 11}. As expected, $\kappa$ becomes small as $\rho\sigma$ increases toward the small 
correlation length regime $\rho\sigma\gtrsim 0.7$. We note that the statistical uncertainty of our 
Monte-Carlo results is about $\pm 0.1$, so that values of $\kappa$ smaller than $0.1$ are not 
statistically distinguishable from zero. On the other hand, very small positive values and slightly negative values 
of $\kappa$ can be randomly obtained, which cannot be conveniently plotted on a log-log graph. 
For this reason, figure \ref{fig: 11} actually shows $\kappa+0.01$, allowing it to keep most of the 
data points inside the viewing range. Thus, for $\rho\sigma\gtrsim 0.7$, our numerical results agree 
with the prediction that the Chern number distribution is Gaussian, and its variance is given by \eqref{eq: 3.2}.

As expected these predictions are not valid when $\rho\sigma$ is small 
(in practice, when $\rho\sigma\lesssim0.7$): Figures \ref{fig: 1} and 
\ref{fig: 11} show that in this regime the scaled Chern variance $\langle\tilde N_n^2\rangle$ is an 
increasing function of $\rho\sigma$, and $\kappa$ is positive. Nevertheless, the data collapse seen 
in these figures is consistent with extended \emph{universality}, where $\langle\tilde N_n^2\rangle$, 
$\kappa$, and plausibly the entire Chern number distribution depends on $\rho\sigma$, but not on 
any other detail of the random matrix element distribution. We conjecture that the universal distribution 
is an exact asymptotic for matrix size $M\to\infty$, but the numerical calculation indicate that it is a 
good approximation already for moderate $M$.

As a last observation on single-band Chern number statistics, we note that while the dependence 
of the scaled Chern number variance on $\rho\sigma$ has no apparent structure 
for $\rho\sigma\lesssim0.7$, the kurtosis follows the power law
\begin{equation}
\kappa\sim (\ell_\kappa\rho\sigma)^{-\alpha}\ ,
\end{equation}
where $\alpha=3.66\pm0.05$, $\ell_\kappa=2.49\pm0.05$ are universal constants.

\begin{figure}[t]
\centering
\includegraphics[width=12.5cm]{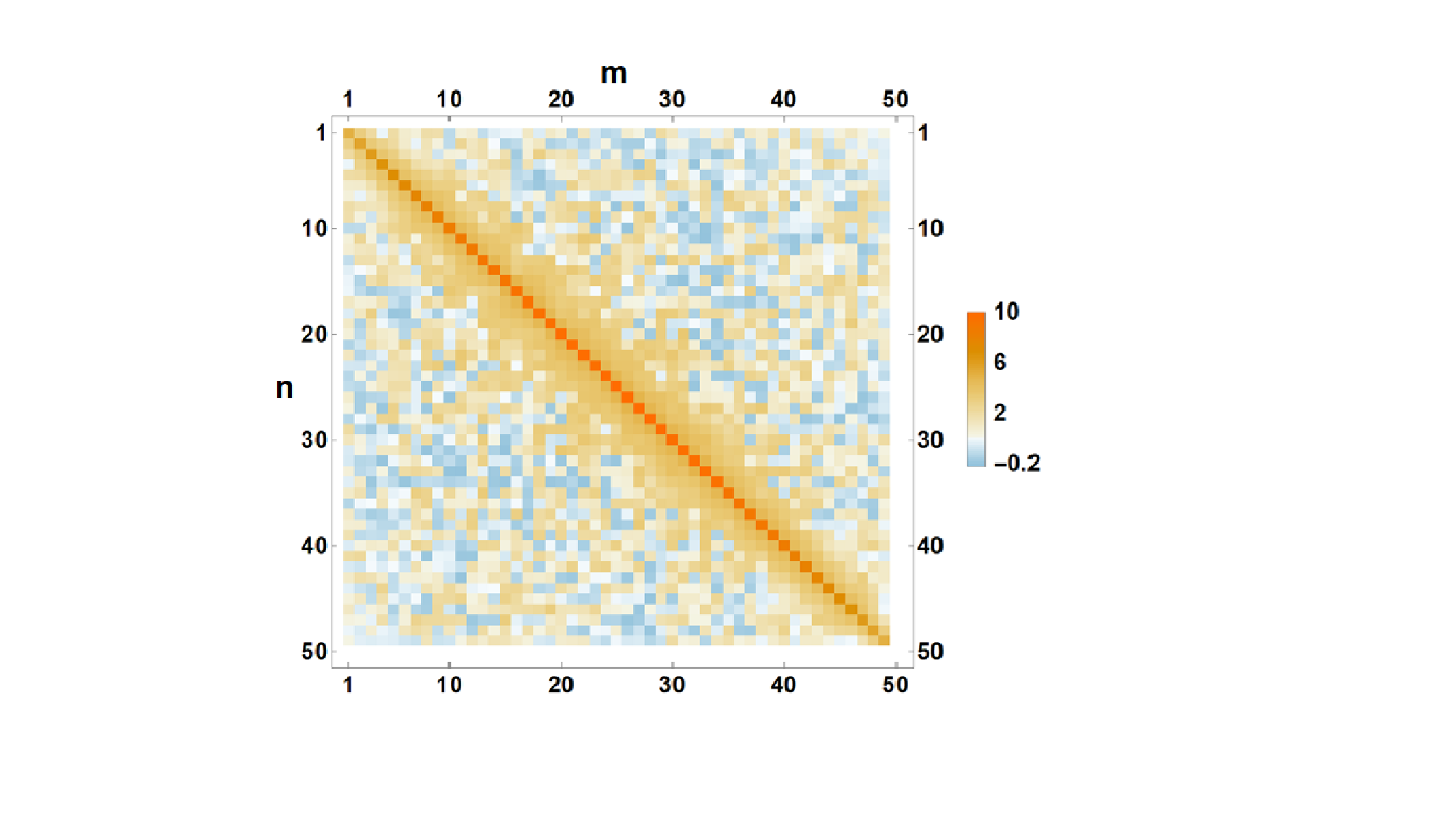}
\caption{
\label{fig: 4}
Plot of the covariance matrix $\langle G_nG_m\rangle$ of the gap Chern numbers of {$50\times 50$} 
matrix fields with a Gaussian matrix-element correlation function with correlation length $r=1$, with values 
coloured according to the legend on the right. It is evident that the diagonal elements of 
the covariance matrix are much larger than the off-diagonal, but the main diagonal is surrounded 
by a thick diagonal band of positive correlations, beyond which the measured correlation are 
comparable with the statistical uncertainty, and are therefore consistent with zero.}
\end{figure}

\section{Gap Chern number correlations}
\label{sec: 5}

{In paper I it was argued that (\ref{eq: 3.7}) is a plausible expression for correlations between 
Chern integers. It was shown that (\ref{eq: 3.7}) is compatible with the 
$\epsilon^{-3}$ scaling of (\ref{eq: 3.5}), 
and this expression appears to be the simplest hypothesis which is compatible with (\ref{eq: 3.5}). 
However, we find that the numerical calculated correlation of the curvature 
of neighbouring bands has a small but 
statistically significant violation of \eqref{eq: 3.7}. The assumption \eqref{eq: 3.7} is 
consistent with the 
hypothesis that the gap Chern numbers are statistically independent.

Here we further test this hypothesis by calculating the gap Chern number 
correlations directly. Figure \ref{fig: 4} 
shows a colormap of the covariance $\langle G_nG_m\rangle$ in a 
population of $50\times50$} matrix fields. 
Evidently, the covariance is positive in a thick diagonal band, even though the off-diagonal terms are much 
smaller than the diagonal terms of the covariance matrix.

The gap Chern number correlations was studied systematically using the Pearson correlation 
\begin{equation}
\label{eq:pearson}
g_{n,k}=\frac{\langle G_{n}G_{n-k}\rangle}{\sqrt{\langle G_{n}^2\rangle\langle G_{n-k}^2\rangle}}\ .
\end{equation}
Our numerical results indicate that the correlation coefficients are independent of $n$ 
(except at the edges of the spectrum), in accord with the homogeneous 
structure of random matrix spectra. We find that the correlation coefficient 
has a universal power law dependence
\begin{equation}
\label{eq:gapc}
g_{n,k}\sim\frac{g_1}{k^{\gamma}} \ ,
\end{equation}
with $g_1=0.18\pm 0.01$, $\gamma=0.62\pm0.01$, for all gap Chern number in the short-correlations 
regime $\rho\sigma\gtrsim0.7$ (see Figure \ref{fig: 5}).

\begin{figure}[t]
\centering
\includegraphics[width=12.5cm]{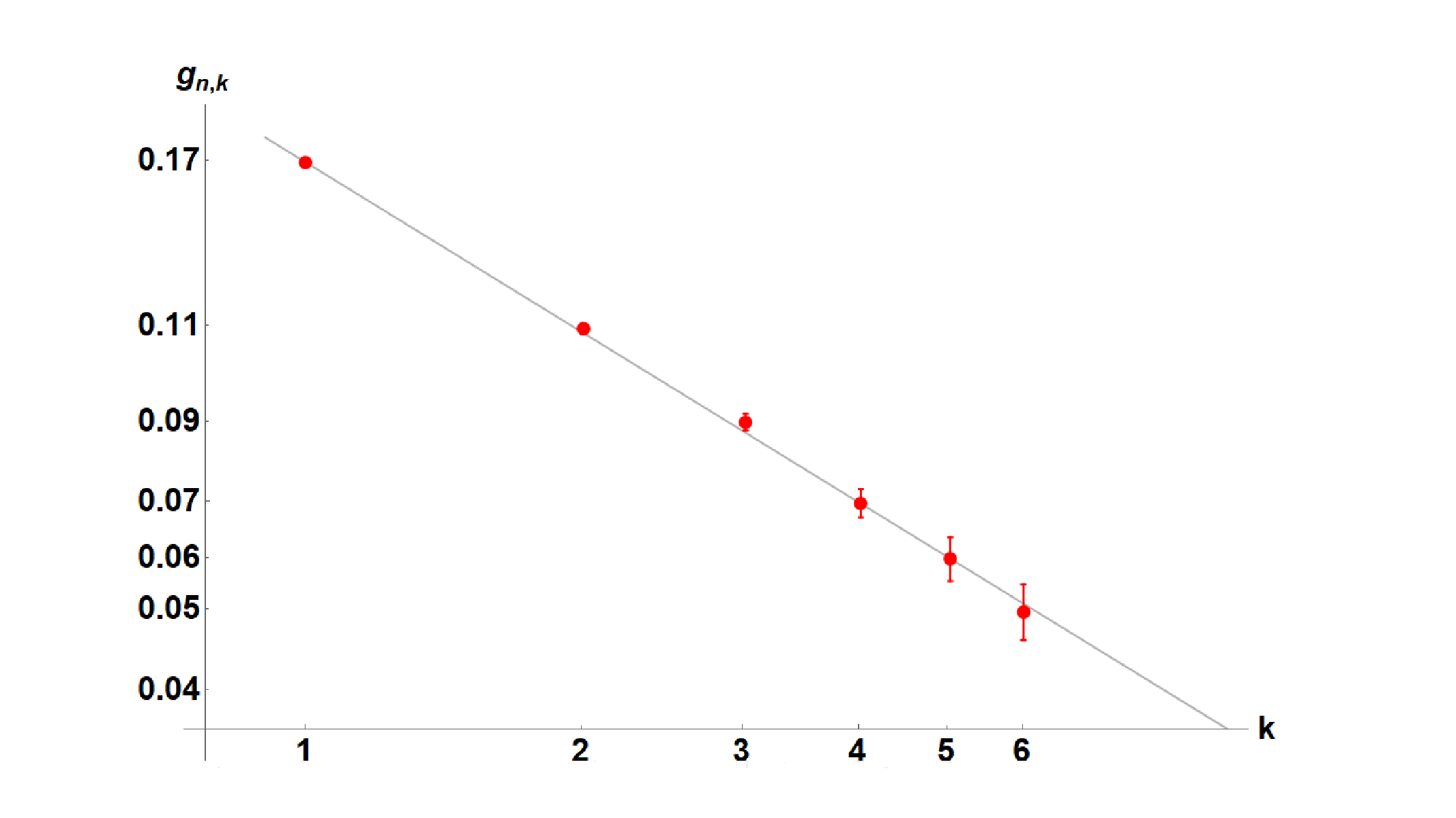}
\caption{
\label{fig: 5}
The off-diagonal correlation coefficient $g_{n,k}$, calculated from the gap Chern number distributions 
of all random matrix sizes and matrix-element correlations in the short correlations regime $\rho\sigma\ge0.7$, 
and all base levels $n$, shown as a function of the offset $k$. The centres and error bars of each 
point show the mean and standard deviation (respectively) of the population of $g_{n,k}$ samples 
obtained in this way; the smallness of the standard deviations supports the hypothesis that the $g_{n,k}$ 
are universal coefficients independent of $n$. The universal function is well described by the power 
law fit $g_{n,k}\sim\frac{g_1}{k^{\gamma}}$ with $g_1=0.18\pm 0.01$, $\gamma=0.62\pm0.01$.}  
\end{figure}

\section{Weighted Chern number variance}
\label{sec: 6}

We computed {the variance of the} weighted Chern number statistic, as defined in 
equation (\ref{eq: 3.6}). We considered a range of values of $\epsilon$, taking $E=0$ and using the largest 
matrices (dimension $M=50$), in both the Gaussian and four-matrix models.
{Figure \ref{fig: 6} shows} the dependence of {$\langle \bar N^2_\epsilon\rangle$} 
on $\epsilon$, using a double-log scale for the Gaussian and four-matrix models, compared 
with values for these variances predicted by  (\ref{eq: 4.2}) and (\ref{eq: 4.3}), and with the asymptotic 
formula (\ref{eq: 4.4}),} on the basis of the independent gap Chern number hypothesis {\eqref{eq: 3.9}.

\begin{figure}[!tbp]
  \centering
  \begin{minipage}[b]{0.49\textwidth}
    \includegraphics[width=\textwidth]{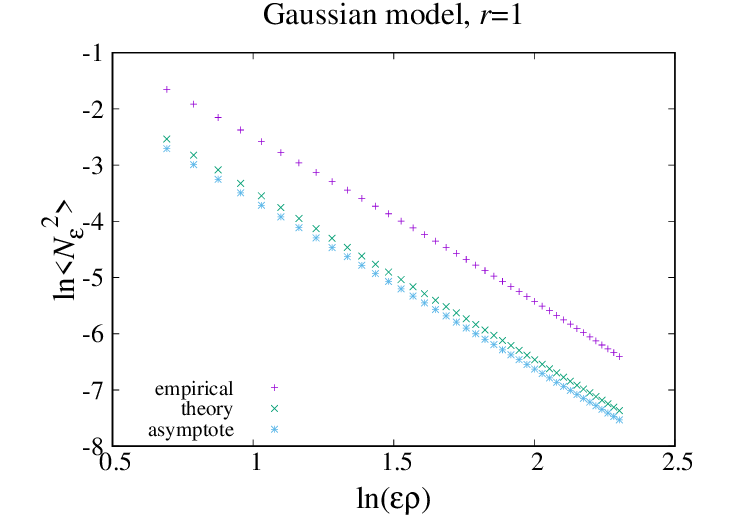}
  \end{minipage}
  \hfill
  \begin{minipage}[b]{0.49\textwidth}
    \includegraphics[width=\textwidth]{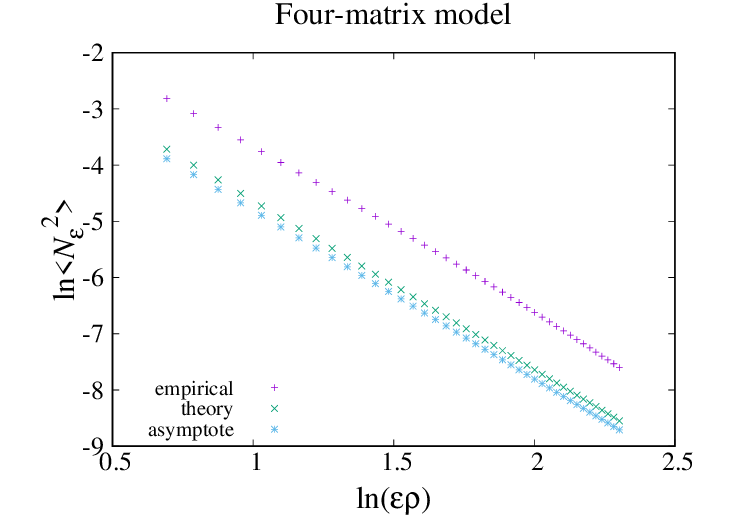}
  \end{minipage}
      \caption{\label{fig: 6}
Double logarithmic plot of the weight Chern number variance $\langle \bar N_\epsilon^2\rangle$ versus the 
product of the density of states$\rho$ and the with $\epsilon$ of the energy window. The purple $+$ symbols 
show $\langle \bar N_\epsilon^2\rangle$ calculated from Monte-Carlo simulation of the random matrix 
model, with energy window {centred} at $E=0$. The green $\times$ and blue $*$ symbols show 
the values of $\langle \bar N_\epsilon^2\rangle$ expected on the basis of independent gap Chern 
number hypothesis, formula \eqref{eq: 4.2}, and its asymptotic approximation \eqref{eq: 4.4}, 
respectively. When $\epsilon\rho$ is large, \eqref{eq: 4.2} agrees with \eqref{eq: 4.4} as 
expected, but both are significantly smaller than the numerical results.}
\end{figure}

Since the density of states is large in our calculations, the values based on {(\ref{eq: 4.2}) and (\ref{eq: 4.3}) are 
well-approximated by (\ref{eq: 4.4})}. However, these predictions, based on the independent gap Chern number 
hypothesis, are not in agreement with the Monte-Carlo results. While (\ref{eq: 4.4})  gives 
the correct order of magnitude of the weighted Chern number variance, the numerically evaluated 
results are greater by a factor as large as three: figure \ref{fig: 8} shows the ratio of the 
numerically obtained $\langle \bar N^2_\epsilon\rangle$ divided by the prediction based on 
\eqref{eq: 4.2}, \eqref{eq: 4.3}, for both the Gaussian and the four-matrix models. 
While we do not have a theoretical explanation for the behaviour of the ratio, the close agreement 
between the two models supports the universality hypothesis for the weighted Chern statistic as well.

\begin{figure}
\begin{center}{
\includegraphics[width=0.65\textwidth]{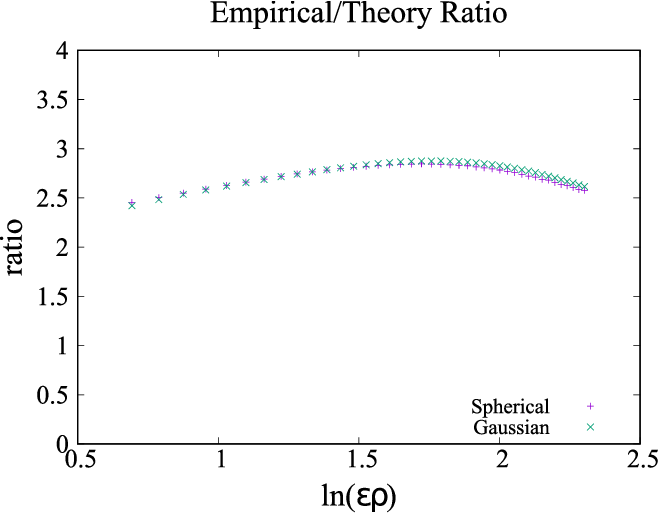}
}
\caption{\label{fig: 8}
Values  of $\langle \bar N_\epsilon^2\rangle$ numerically calculated for the four-matrix model 
(purple $+$ symbols) and Gaussian correlations model (green $\times$) divided by the prediction 
based on formula \eqref{eq: 4.2} and \eqref{eq: 4.3}, shown as a function 
of $\ln (\rho\epsilon)$. The significant deviation 
of this quotient from one is a consequence of the failure of the independent gap Chern number hypothesis, 
which spoils the strong cancellation in \eqref{eq: 4.2} and \eqref{eq: 4.3}. 
The close agreement between the ratios obtained for 
the two random matrix models is consistent with the Chern-number-statistics universality hypothesis.}
\end{center}
\end{figure}

It is interesting that while the gap Chern number correlations studied in section \ref{sec: 5} are small, 
the deviations from (\ref{eq: 4.2}) and (\ref{eq: 4.3}) are quite significant. 
This may be an indication that there are non-trivial multi-level correlations between Chern numbers,
but further study of this question is beyond the present scope.

\section{Conclusions}
\label{sec: 7}
The present numerical study of the Chern numbers of parametric random matrix models 
examined three types of statistics: moments of the single-band Chern number distribution, 
correlations of the gap Chern numbers, and the variance of weighted sums of band Chern numbers.

The single-band statistics confirmed that when the parametric correlation length is small, 
the band Chern number distribution approaches a Gaussian, and that 
the proportionality factor is universal, 
consistent with the results of paper I \cite{Gat+21}. It showed furthermore that the distribution is 
non-Gaussian but universal in the previously unexplored long correlations regime, and that the 
kurtosis of the distribution depends as a power law on the correlation length in this regime, 
for which we have no theoretical explanation at this point.  Note also that in 
the limit where the correlation length tends to infinity, the probability that Chern number 
is nonzero tends to zero, and then the kurtosis can be estimated on the basis of this 
probability; however, the correlation lengths that we study here are not large enough 
to make this estimate valid.

We found that the gap Chern numbers are correlated, refuting an earlier hypothesis. 
The gap Chern  number correlations are weak, but decay slowly as a power law when the 
spectral separation between the gaps increases. Unlike the band Chern numbers 
that are strongly anticorrelated, there is no compelling reason for correlations between 
the gap Chern numbers. Since the gap Chern numbers on the sphere can be expressed 
as the signed number of band intersections in the sphere's interior, an analysis of the 
correlation between the intersections should be able to elucidate the cause of the gap 
Chern correlations.

When we looked at the variance of a weighted sum of Chern numbers, however, 
we found that there was a very significant difference from the predictions based upon 
assuming independent gap Chern numbers.
One of the motivations for looking at statistics of Chern numbers was the 
hypothesis that there are cancellation effects, which reduce the variance of 
a sum of Chern numbers. We found that the variance of the weighted sum was 
\emph{larger} than the prediction in paper I, indicating that the cancellation effects 
are weaker than anticipated.

We were motivated to perform this study by the success of random matrix models 
as exemplars of universal properties of complex quantum systems. Our results do show 
how random matrix models for Chern integers can be successfully quantified within
the framework of a \lq universality' hypothesis. However, it is not clear to what extent 
these random matrix models are representative of the quantised Hall effect in 
physically realistic models. In particular, when the dimension of the random matrix
is large, then sum of the band Chern numbers may be a very large number. 
While we have shown that there is significant anti-correlation between the Chern 
integers of adjacent levels, we have not been able to show that the sums of Chern numbers 
for these random matrix models are compatible with predictions from semiclassical 
theories. One possibility is that the sums of Chern numbers for physical systems 
show a greater degree of cancellation than those of random matrix models. Another 
possibility is that the bands of generic complex systems overlap, so that the Fermi 
energy cannot lie in a gap. In this latter case, the argument that the Hall conductance 
can be expressed as a sum of Chern integers fails. 
It is, therefore, desirable to complement this present work with studies of Chern numbers 
in physically realisable models for complex quantum systems.

\section*{Acknowledgements}
MW is grateful for the generous support of the 
Racah Institute, who funded a visit to Israel. 

% TODO: include author contributions
%\paragraph{Author contributions}
%This is optional. If desired, contributions should be succinctly described in a single short paragraph, using author initials.

% TODO: include funding information
\paragraph{Funding information}
This work was supported by the German-Israeli Foundation for financial support under grant number GIF I-1499-303.7/2019.
% Correctly-provided data will be linked to funders listed in the 
%\href{https://www.crossref.org/services/funder-registry/}{\sf Fundref registry}

\appendix
\section{Monte Carlo simulation of Chern numbers}\label{app:mc}
Our numerical calculation of the Chern number distribution is based on Monte-Carlo 
sampling of the parametric GUE random field on the two-sphere. Each matrix element 
field realisation is defined by a set of random spherical harmonics amplitudes $a_{jk,lm}$
\begin{equation}\label{eq:hjkmc}
H_{jk}(\textit{\textbf{p}})=\sum_{jk,lm}a_{jk,lm}Y_{lm}(\textit{\textbf{p}})\ ,
\end{equation}
$Y$ being standard spherical harmonics. The amplitudes $a_{jk,lm}$, $a_{j'k',l'm'}$ 
are statistically independent unless $j=j',k=k'$ or $j=k',k=j'$, and $l=l',m=m'$ and their variance is 
\begin{equation}
\langle|a_{jk,lm}|^2\rangle=c_l\ ,
\end{equation}
where $c_l$ are the coefficients of the expansion of the matrix-element correlation 
function in Legendre polynomials.

The Chern number of a given matrix field realization is calculated using the algorithm of 
Fukui {\sl et al.} \cite{Fuk+05} adapted to a triangulation of the two-sphere. The matrices 
\eqref{eq:hjkmc} are calculated at the vertices of the triangulation, diagonalised, 
and the eigenvectors are used to estimate the the flux of the adiabatic curvature through 
each triangle, as explained in \cite{Fuk+05}. The Chern number is then estimated by the 
sum of the adiabatic curvature flux of the entire triangulation.

Even though the Chern number obtained by this algorithm is guaranteed to be an 
integer, avoided crossings may cause it to produce an erroneous result if the triangulation 
is too coarse. For this reason we used an adaptive mesh algorithm, where starting from an 
octahedral triangulation, each triangle was subdivided into four congruent equilateral 
triangles; the total curvature flux of the four sub-triangles was compared to that of the 
original triangle, and the result was accepted if the difference between the two flux 
calculation was smaller than a numerical tolerance parameter. Otherwise, the 
triangulation was further refined until convergence was achieved. The nonuniform 
triangulations produced by this method were essential for numerical tractability of 
the calculation.

The numerical codes are available at:\\ {\tt 
https://github.com/orswartzberg/Numerical-calculation-of-Chern-numbers}.

\end{document}